# E-Quarantine: A Smart Health System for Monitoring Coronavirus Patients for Remotely Quarantine

Doaa Mohey El-Din[1,*], Aboul Ella Hassanein[1,*], Ehab E. Hassanien[1], and Walaa M.E. Hussein[2]

[1]Faculty of Computers and Artificial Intelligence, Cairo University, Egypt
[2]Clinical Pathology, Faculty of Medicine, Cairo University, Egypt
[*]Scientific Research Group in Egypt (SRGE)
d.mohey@alumni.fci-cu.edu.eg, aboitcairo@cu.edu.eg, E.Ezat@fci-cu.edu.eg
w.mohy@pg.cu.edu.eg

*Abstract*— coronavirus becomes officially a global pandemic due to the speed spreading off in various countries. An increasing number of infected with this disease causes the Inability problem to fully care in hospitals and afflict many doctors and nurses inside the hospitals. This paper proposes a smart health system that monitors the patients holding the Coronavirus remotely. Due to protect the lives of the health services members (like physicians and nurses) from infection. This smart system observes the people with this disease based on putting many sensors to record many features of their patients in every second. These parameters include measuring the patient's temperature, respiratory rate, pulse rate, blood pressure, and time. The proposed system saves lives and improves making decisions in dangerous cases. It proposes using artificial intelligence and Internet-of-things to make remotely quarantine and develop decisions in various situations. It provides monitoring patients remotely and guarantees giving patients medicines and getting complete health care without anyone getting sick with this disease. It targets two people's slides the most serious medical conditions and infection and the lowest serious medical conditions in their houses. Observing in hospitals for the most serious medical cases that cause infection in thousands of healthcare members so there is a big need to uses it. Other less serious patients slide, this system enables physicians to monitor patients and get the healthcare from patient's houses to save places for the critical cases in hospitals.

## I. INTRODUCTION

Recently, coronavirus (COVID 19) becomes a pandemic virus that patients reach more than 1.856.798 around the world [1]. In April 2020, the death numbers from this virus reach 114.312 around the world [1, 2]. The percentage of deaths achieves more than 18% daily. The spreading of infected people and deaths numbers are increasing daily. Most people are affected by patients are healthcare members whether doctors or nurses. The recent statistics of the infected healthcare members in doctors and nurses reach more than 65.000 [3]. Thousands of them die from the infection of patients holding coronavirus in serving healthcare in hospitals.

So, there is a big need to go forward the remote healthcare especially in the highly infected viruses such as COVID 19 to save lives. Remotely healthcare monitoring requires multiple sensors to record the parameter of each case in real-time to improve the healthcare services in speed time and decision making remotely [4]. The smart healthcare system relies on the integration between artificial intelligence and internet-of-things technologies.

Previous research target are monitoring the patient in various diseases such as debates, diets, and after surgical operations. They enable physicians to observe multiple patients at the same time. That eases the system to be highly flexible, and accurate. The used sensors are variant types whether wearable or built-in sensors or mobile sensors [5]. These systems require interpreting the extracted data from these sensors to reach the main objective.

This paper presents the EQuarantine system that is a proposed smart Health System for monitoring coronavirus patients for remote quarantine. It becomes important to save thousands of lives from infection or death. It is based on fused multiple data from various sensors to detect the degree of development of the disease and the seriousness of the health condition. It is based on monitoring the readings' heart pulse, respiratory rate, blood pressure, Blood PH level in real-time. It proposes to be a time-series data that includes sequenced data points in a time domain. The data extracted from multiple sensors are gathered sequentially based on multi-variable measurements. It proposes a classification of patient's cases. It also targets observing multiple users concurrently. The proposed technique will be constructed based on the combination of fusion types, feature level, and decision level. It will rely on the long-short term memory (LSTM) technique that is considered a deep neural network (DNN) technique for sequenced data. It uses the power of feature learning ability and improves the classification of serious health condition levels. Then using the Dempster-Shafer fusion technique for fusion decision.

The proposed system enables monitoring patients from their homes that save governmental cost and time through measuring the changes in patient's medical readings. It will serve humanity in the reduction of Coronavirus infection and save healthcare members around the world. It also saves hospital places for emergency cases.

The rest of the paper is organized as the following: Section II, related works, section III, the dangers of Coronavirus, section IV, the proposed smart health system, section V, experiments and results. Finally, section VI targets the conclusion outlines and future works.

## II. RELATED WORKS AND BACKGROUND

Several motivations research and investment in the smart health systems or telehealth systems that are simulated real system for observing or following diseases remotely whether in hospitals or patient's homes [6, 7, 8, 9, 10, 11]. These motivations target saving lives, time and cost. The main objective of smart health is remotely controlling for many patients and monitoring their diseases follow in real-time.

The essential challenge in smart health is interpreting, fusing and visualizing big data extracted from multiple smart devices or sensors. It improves the making of decision quality for medical systems. The data are collected from sensors to observe patients remotely at their houses. These data are applied by the aggregation and statistical ways for the decisions of the medical system.

### a. Smart health

Smart health is a hot area of research and industry which includes a connection between sensors deals with the patients. It can provide monitoring of the remote patient through several ways as video, audio or text. The main problem of this area how to manage the data, analytics them and visualize the reports of the data. This section presents a comparative study of several previous researches on smart health. It includes a combination of artificial intelligence and machine learning algorithms that will support high results of prediction and evaluation of the patient's problems.

Previous researches present several smart health motivations for constructing a suitable system for observing patients based on each medical case as shown in Table (1). Researchers in [6], build graphical smart health system for visualizing patient's data for remotely physicians. Noisy data and redundant features become the main limitations are faced with this system. Researchers in [7], present a new smart health system with high accuracy for observing patients after surgeries. This system requires medical experts and huge analysis from doctors to support the full vision for each case. Researchers in [8], improves the monitoring patients remotely with accuracy 9%. They face several limitations in reliability and integrity. Researchers in [9, 10, 11], build smart medical system for hospitals to be powerful in monitoring patient's cases. However, it still requires motivation for enhancing accuracy.

Table. 1: a comparative study of smart health systems for monitoring patients

| Paper No. | Domain | Target | benefits | Limitations |
|---|---|---|---|---|
| [6] | Smart Health | Monitoring patients remotely | Visualize patient cases graphically | Noisy data and redundant features |
| [7] | Smart Health | It is based on creating surgical prediction multi-model for patients | 95% accuracy results | Lack of information (requires to expert people) |
| [8] | Smart Health | Support monitoring healthcare | İmprove accuracy 9% | Improving reliability and integrity |
| [9] | Smart Medical | Improves the hospital recommendations | Enhance patient monitoring | Requires enhancing the accuracy |
| [10] | Smart Health | Monitoring patients | Improves patient monitoring | Requires improving accuracy |
| [11] | Smart Health | Monitoring patients | recognize 4 parts: falls, lying, standing, sitting and walking activities | The hardness of fusion with various data types |

From previous motivations, finding to construct any smart health system requires to know all conditions, and some expert knowledge to make observing automatically and detect the important readings or anomalies for each patient. That requires supervised training to support any new cases and detect problems. Data visualization for the extracted patient's data is very important to save time and lives simultaneously. Data visualization refers to one of the main fields of Big Data analytics that enables end-users to analyze, understand, and extract insights.

### b. Deep learning technology

The essential idea of deep learning depends on the artificial neural networks (ANNs) study [12]. ANNs are a new trend for the active research fields due to building a standard neural network (NN). It uses neurons for producing real-valued activations and, by adjusting the weights, the NNs behave as expected. The approaches of deep learning have been utilized powerfully in big data analysis for various applications and targets [13]. They use for analyzing the computer vision, pattern recognition, speech recognition, natural language processing, and recommendation systems. There is a trade-off between the accuracy measurements and the complexity when applying the deep learning algorithms. These approaches include several types as a convolutional neural network (CNN), recurrent neural network (RNN), and [14]. Long Short-Term Memory Recurrent Neural Networks (LSTM-RNN) is one of the most powerful dynamic classifiers publicly known.

### c. Data fusion process

It is considered a framework where data from multiple sources are gathered, mixed aggregated to make them more powerful and more adapted to a given application. Data fusion means the process to reach higher efficiency results to deal with multiple and heterogeneous data sources. There are different types of patient-doctor communication video, audio or text message through remote IoT devices or mobile sensors for computing bio-physical characteristics. In smart health systems, a system is responsible for measuring the effective requirement of therapy or other health-related issues only considering interviews and data gathered at the patient home.

Table.2: A comparative study between research fusion levels

| Paper No. | Fusion Type | Target | Benefit | Limitation |
|---|---|---|---|---|
| [16] | Feature | Predicting the stock market behavior with the aid of independent component analysis, canonical correlation analysis, and a support vector | High accuracy in prediction analysis | Requires motivation for improving the parallel fusion |
| [17] | Feature | an automated feature construction approach | search method for finding a set of | Requires improving in extracting other features |

|  |  |  | relevant features |  |
|---|---|---|---|---|
| [18] | Feature | Investigate the performance of the sparse autoencoders utilized in regression analysis. | Improve accuracy | Improving performance |
| [19] | Decision | the local weighted linear prediction algorithm | reach desirable results based on training synthetic ECG signals. | Improve performance |
| [20] | Decision | Multiple decision trees weighting fusion algorithms | Enhance the ability of information abstraction. | Improve performance |
| [21] | Hybrid | Targets adaptive fusion model | Better than unimodal fusion in terms of accurate results. | Requires to be adaptive in multiple domains |
| [22] | hybrid | a hybrid fusion strategy is used for the feature and score level | Improves accuracy for recognition 98.7% | Raising research in multi-biometrics. |

From the previous motivations (refer to Table (2)), finding the hybrid between feature level and decision level is more complex however, it has better accuracy.

### III. THE DANGEROUS OF CORONAVIRUS

COVID-19 is inherited from previous viruses (as SARS) but the evolution of this virus becomes more dangerous and higher spreading. Although COVID-19 is less dangerous than SARS on most people, it is a higher infection between people. Compared with other viruses in 2003 and 2012 [23, 24], they are limited to some areas and took limited time. The research motivations predict each new COVID-19 case infects 2.5 other people on average when no effort is made to keep people apart [25]. Coronaviruses are jumping increasingly from animals to humans, creating new threats.

Table.3: a comparative study between COVID-19 and Sars Viruses

|  | Conrona (Covid-19) | Sars (SARS-CoV) |
| --- | --- | --- |
| Disease type | Pandemic | Not pandemic |
| Danger | killed more than 3 percent of confirmed cases | Sars killed 10 percent of infected individuals |
| Affected by | respiratory disease | Viral respiratory disease |
| Spreading | High spreading | low spreading |
| Infected People | High infection | Lower infection |
| Infected countries | More than 181 country | Around 27 country |
| Dangerous on people's slide | Low immunity especially, pregnant and old people. | Low immunity especially, pregnant and old people |
| Number of infected people | More than a million in three months (1.856.798) | 75.000 |
| Number of deaths | 114.312 | 813 |

The new coronavirus has killed nearly 3 times as many people in 8 weeks as SARS did in 8 months [26]. Table (3) disuses the comparative study between COVID-19 and Sars in several dimensions. The main reason for the dangers of this virus is a bad impact on respiratory disease. That is caused by increasing high-risk cases by 20 percent of patients and killing more than 3 percent of confirmed cases. Sars is caused by killing 10 percent of infected patients. Older people, whose immune defenses have declined with age, as well as those with underlying health conditions, are much more vulnerable than the young [27]. However, death rates are hard to estimate in the early stages of an epidemic and rely on the medical care given to patients. Lack of healthcare equipment cause killing people due to this virus has a bad effect on breath and respiratory rate. For instance, ventilators protect lives by causing pneumonia to breathe.

## IV. THE PROPOSED SMART HEALTH SYSTEM FOR CORONAVIRUS

The proposed smart Health System aims at monitoring coronavirus patients for remotely quarantine. It targets saving thousands of lives from infection or death. It depends on the integration between artificial intelligence and internet-of-things for fusing multiple sensory data from various medical sensors to detect the degree of development of the disease and the seriousness of the health condition. The proposed system improves decision making quickly and simultaneously. Figure (1) shows the big image of the proposed smart health system for monitoring infected coronavirus remotely based on Internet-of-Things devices

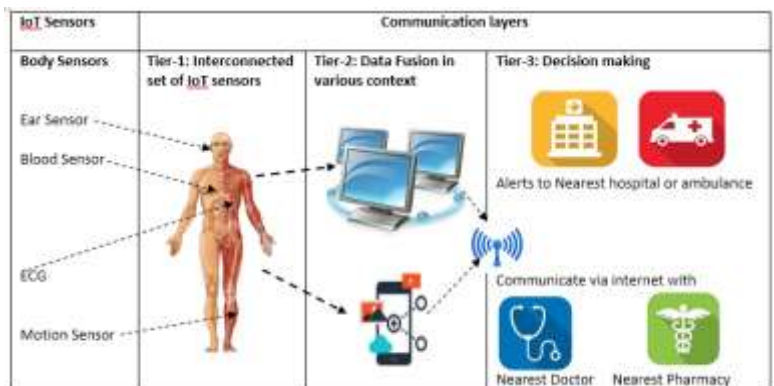

Figure.1 The proposed smart health system for monitoring infected coronavirus remotely based on Internet-of-Things devices

It is based on monitoring the reading's heart pulse, respiratory rate, and blood pressure, Blood PH level in real-time. Figure (1) presents the proposed smart health system based on the communication between IoT devices in a network. This consists of three tiers: tier1 deals with the different sensors connected by the patient such as in mobile, wearable, IoT devices, or accumulators' sensors for measuring the patient evolution case (as blood pressure). Tier 2 the fusion between the data in multi-sources mostly in different multimedia. The tier3 the visualization and deciding for emergency cases, making profiling for each case, and how to predict the next problem for each patient.

It proposes to be a time-series data that includes sequenced data points in a time domain. The data extracted from multiple sensors are gathered sequentially based on multi-variable measurements. It proposes a classification of patient's cases. It also targets observing multiple users concurrently. The proposed technique will be constructed based on the combination between the fusion feature level and fusion decision level. It will rely on the long-short term memory (LSTM) technique that is considered a deep neural network (DNN) technique for sequenced data [28]. It uses the power of feature learning ability and improves classification serious health condition level. Then using the Dempster-Shafer fusion technique for fusion decision [29].

### a. Architecture life cycle

The proposed system enables monitoring patients from their homes that save governmental cost and time through measuring the changs in the patient's medical readings. It will serve humanity in the reduction of Coronavirus infection and save healthcare members around the world. It also saves hospital places for emergency cases. The life cycle of smart health technique includes six main layers, cleaning data layer, anomaly detection layer, extracting features based on deep learning, LSTM deep learning layer, and fusion layer as shown in the lifecycle in Figure (2).

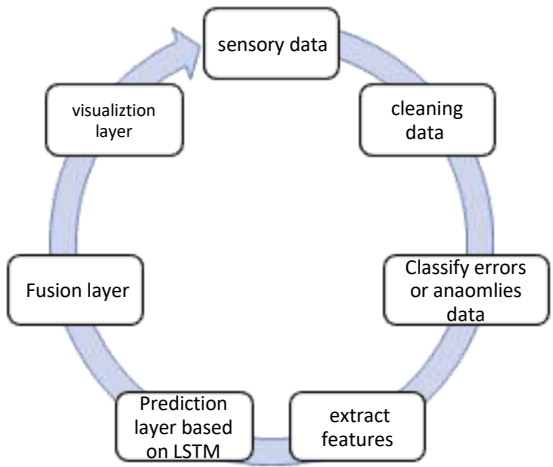

Figure 2: The lifecycle architecture of the proposed system

The architecture of the smart health technique consists of the main six layers, cleaning data, feature extraction layer, deep learning techniques, fusion layer a shown in Figure (2).

The first layer, cleaning the data layer uses for pre-processing time-series data. the second layer includes anomaly detection controlling for observing and recording any changes in the data such as outliers or errors. The importance of this layer appears in neglecting the context or domain and focusing on the main three dimensions data types, major features, and anomalies to unify the technique in any domain. Third layer, the feature extraction layer refers to the automated extracting features from any context. However, not all features are important and require fusing for the target. so, that requires feature reduction or ignorance of some features before the fusion process. The fourth layer, Deep learning algorithm that is based on the input data type. The deep neural model is constructed based on target and features. The fifth layer, the Fusion layer refers to fuse multiple data from multiple sensors. This layer manages the fusion process based on three dimensions the input data types, features, and anomalies in input data.

### b. The hybrid health system Algorithm

The proposed algorithm is a hybrid fusion technique between feature fusion level and decision fusion level concerning determining anomalies to improve decision making quickly and simultaneously. The hybrid technique is illustrated in making a decision based on extracting new features from the data and fusing the decisions from tracing in each sensor. The proposed system enables to classify patients in the risk level and make decisions concurrently. It also predicts each patient's evolution case based on a remote monitoring process.
The algorithm includes the main six layers.

**(1)** Cleaning data layer: The nature of time-series data usually includes several noisy data or missing data. So, this layer targets ensuring the quality of the input data that ignoring missing data, determining error readings, filtering anomalies or fixing structural errors. It also determines the duplicate observations and ignores irrelevant notifications. This layer is very important to make the system highly reliable. The steps of this layer are:
- Check on empty records, or terms in each cell.
- Check on duplicate records
- Check on noisy data
- Check on unstructured data to convert the suitable structure for the dataset.

**(2)** Anomaly detection layer: The previous layer can filter anomalies or readings errors [30]. These anomalies mean identification rare events that are happening and affected on other observations. They usually require making decisions quickly and concurrently. This layer classifies anomalies to improve the prediction analysis of patient's cases. Algorithm (1) describes the main steps of anomaly detection.

| Algorithm (1) Anomaly Detection layer |
|---|
| 1.    Int c=0; //column number, <br> 2.    int r=0; // row number <br> 3.    int t=0; // term in each cell <br> 4.    int normal; // that refers to the interval for each cell based on the entire condition <br> 5.    For (int a=0; a< c; c ++) <br>       { <br> 6.    For (int b=0; b< r; r++) <br>       { <br> 7.    If (t ==null) <br> 8.    Delete r; <br> 9.    Else <br>       { <br> 10.  If (t ⊆ normal) <br> 11.  Continue; <br> 12.  Else <br> 13.  { <br> 14.  For (int n=0; $t_{n+1}$; n++) <br> 15.  Check ($t_{n+1}$); // check any change in any cell in the same record to determine the errors or anomalies <br> 16.  If finding no changes that usually will be classified into error in readings <br> 17.  To ensure this classification, compare the row has changed and before or after to check errors <br> 18.  If finding changes or effects on other terms in the row's cells. That will be classified into the anomaly. <br> 19.  // Making anomalies classification for emergency cases. <br> 20.  } } } } |

**(3)** Features level: The feature extraction layer or feature engineering layer refers to the interaction between extraction features [31]. It deletes redundant and unused features. The extracted features are medical and require expert knowledge. The combination of the two features can interpret a new feature for improving results and decision making.

Heart rate and respiratory rate are the significant keys to estimate the physiological state of people in several clinical properties [32, 33]. They are utilized as the main assessment in acutely sick children, as well as in those undergoing more intensive monitoring in high dependency. The heart rate and respiratory rate are important values utilized to detect responses to lifesaving interventions. Heart rate and respiratory rate keep an integral part of the standard clinical estimation of people or children presenting. The emergency cases will be classified based on the outranges of these parameters. The previously computed median of the representative centiles (1st, 10th, 25th, 75th, 90th, 99th) for the data from each included study. Finding each age has a different normal range of heart rate and respiratory rate.

| Algorithm (2): feature extraction layer |
|---|
| 1. Set r0 = mean respiration features; |
| 2. H0 = mean heart features; |
| 3. μt = mean and |
| 4. σt = standard deviation |
| 5. set τ = μt + 2σt. // Threshold of the allowed skewness |
| 6. set Rm = R0 and Hm = H0. // Rm and Hm are the most recent respiration and heart feature vectors |
| 7. set Tr = 0 // the time at which respiration features captured. |
| 8. set Th = 0 // the time at which heart features captured. |
| 9. Fr = T × (H − MH) + MR,<br>T = CF H × C −1 S, it shows how the respiration features are transformed to the coronavirus feature |
| 10. For (int t=1; t++)<br>{ |
| 11. If ( R is available ) |
| 12. Update Rm and TR, |
| 13. Update Hm and TH. |
| 14. If (T-TR) > t |
| 15. Then |
| 16. Set Rm=R0. |
| 17. If (T-Ts > t) |

> 18. Then
> 19. Set Hm=H0.
> 20. Set Rbm =M ap (Rm)
> 21. Classify features vectors.
>     }
> 22. End for

**(4)** Predication Layer for each patient's case evolution: Artificial intelligent technique is used for monitoring patients and their case evolution. Long-short term memory (LSTM) is a deep learning technique for sequenced data [32]. This layer is based on LSTM to predict the future case for each patient based on previous disease readings.

**(5)** Fusion layer: There are several fusion techniques but this research applies the Dempster-Shafer fusion [33]. Fusion process contains the following steps:

> **Algorithm (3): Dempster-Shafer Fusion Technique**
>
> 1. Let $\Theta$ refers to a frame of discernment. It means a group of all mutually and exhaustive propositions. Let $2\Theta$ express the power set of $\Theta$. For each proposition in $2^\Theta$, a probability mass m is assigned subject to the rules that
> $$m(\phi) = 0$$
> $$\sum_{A_i \in 2^\Theta} m(A_j) = 1$$
>
> 2. The belief or support function refers to a lower bound on the probability of proposition A and is known as
> $$Spt_i(A) = \sum_{A_j \subseteq A} m_i(A_j)$$
>
> 3. The plausibility function refers to an upper bound on the probability of proposition A and is known as
> $$Pls(A) = 1 - Spt(\sim A)$$
>
> 4. The uncertainty interval of proposition A is [spt(A), pls(A)] and the uncertainty of proposition A is given by
> $$u(A) = Pls(A) - Spt(A)$$
>
> 5. For each possible proposition (e.g., user-A), Dempster-Shafer theory grants a rule of gathering sensor Si's supervision mi and sensor Sj's

> observation mj
>
> $$(m_i \oplus m_j)(A) = \frac{\sum_{(E_k \cap E_{k'})=A} m_i(E_k) m_j(E_{K'})}{\sum_{(E_k \cap E_{k'})=\emptyset} m_i(E_k) m_j(E_{K'})}$$
>
> This combining rule can be generalized by iteration: if we treat mj not as sensor Sj's notification, but rather as the already fused by Dempster-Shafer theory notification of sensor Sk and sensor S.

**(6)** Visualization layer: The Fusion output constructs a full vision for patient cases based on extracted sensory data for each patient. The physicians may be finding a hardness in readings hundreds of patients concurrently. So, visualizing data is a very important layer to classify the serious condition level for patients (high risk, medium, low) that may use colors (for example, red color for high-risk cases, yellow for medium risk cases, green for the normal rate or low-risk cases) to getting attention from doctors quickly to improving decision making simultaneously and quickly in Figure (3).

**(7)** Output: That includes colorized and classified data for all patients for each observer doctor. It also includes a detailed sheet for the evolution of each patient and the prediction risk of their patients.

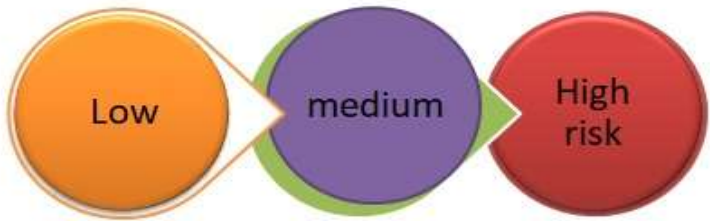

Figure.3: The patient cases classification into one from the three classifications (Low risk, medium risk, and high risk)

## V. EXPERIMENT & RESULTS

The recent governmental motivations go forward to reach high benefits from smart health. It also starts collecting real data concurrently for saving health members' lives and saving patients.

### a. Dataset

The proposed system targets multiple sensors or IoT devices to record patients in quarantine in Figure (4). It improves healthcare working due to the lack of healthcare equipment and reduces infection between healthcare members. The proposed fusion technique for the real dataset that covers the real datasets, the input dataset of Coronavirus quarantine is a modified dataset from the input data as shown in Table (4).

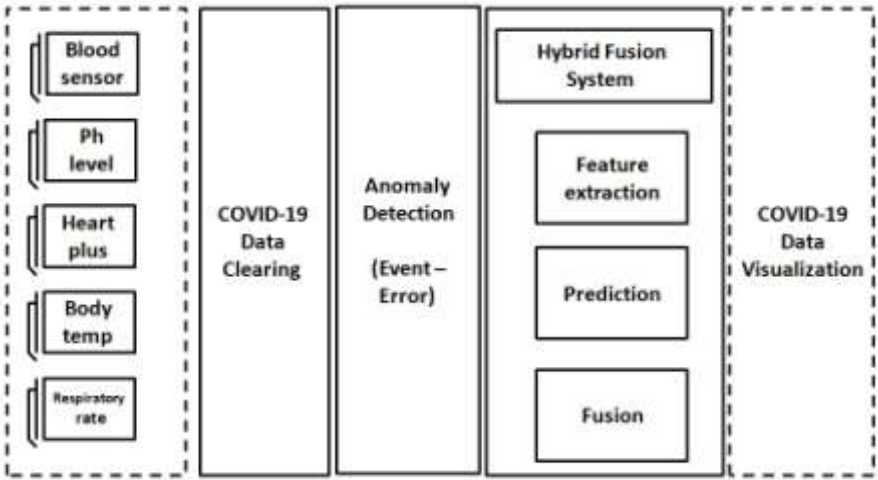

Figure.4: The proposed Smart Health System

The training dataset is based on the collected several real datasets for normal and patients. The fusion process is very critical between multiple sensory data to reach the full vision of the corona disease level. The fusion technique depends on fusing many features of coronavirus that includes heart pulse, respiratory rate, body temperature, blood pressure, and Blood PH level in real-time. They are the main features in fusion process in Figure (5). The fusion technique targets improving the prediction of the disease features and protecting patients' and healthcare members' lives.

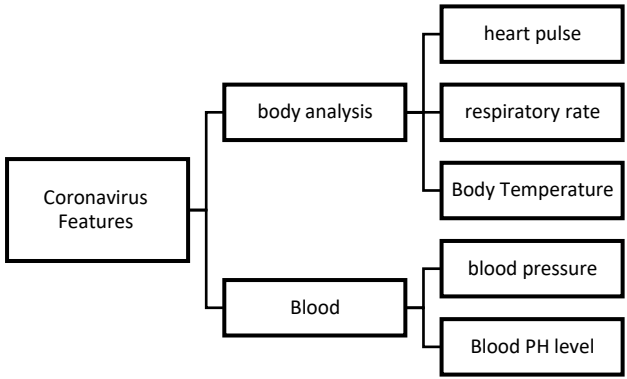

Figure 5: The Essential Recording Features Coronavirus

Table.4: training dataset is collected from two real separated datasets

| Dataset | Ref | Description | Dataset size |
|---|---|---|---|
| Cardiovascular Disease dataset | [34] | Including 11 properties About Patients (it includes patient's profile age, height, weight, gender) with a medical profile (systolic blood pressure, diastolic blood pressure, cholesterol, glucose, smoking, alcohol intake, physical activity, presence or absence of cardiovascular disease) | 70 000 records of patient's data |
| Respiratory Sound Database | [35, 36] | the annotation text files have four columns: Beginning of respiratory cycle(s), End of respiratory cycle(s), Presence/absence of crackles (presence=1, absence=0), Presence/absence of wheezes (presence=1, absence=0) | 126 Audio for patients |

There is a need to making many modifications to the dataset due to the lack of sensory data for patients (refer to Table (4)) in terms of the real normal ranges for normal people as shown in Table (5). These modifications include two steps, data augmentation and generated time. The data will be enlarged to 700.000 records by observing each patient into the previous 10 hours to support predicting the next case rate and save lives before sudden actions in the "cardiovascular dataset". These data set is fused with the real ranges of Blood PH level and body temperature ranges.

Table.5: Normal ranges for people concerning the age

| Age<br>Medical dimension | <18 | 18-25 | 28-35 | 36-45 | 45-55 | 56-65 | 65+ |
|---|---|---|---|---|---|---|---|
| Heartbeat rate average | 70-73 | 70-73 | 71-74 | 71-75 | 72-76 | 72-75 | 70-73 |
| Blood PH level | \multicolumn{7}{c}{7.35 and 7.45} |
| Body temperature | \multicolumn{7}{c}{36.1°C - 37.2°C} |
| Blood pressure | 120/80 | 120/80 | 134/85 | 137/87 | 142/89 | 144/90 | 144/90 |
| Respiratory rate | 25-35 | 18-20 | \multicolumn{5}{c}{10-22 percentile} |

### b. *Experiments*

The processing technique includes six layers, that are applied as the following.

The proposed E-Quarantine system uses A long short-term memory network (LSTM) twice times, first using for detecting anomalies and for predicting the evolution case of each patient based on previous disease profile. LSTM is considered an evolution from recurrent neural network technique. The LSTM technique is very powerful for classifying sequential data. The most common way to making training on RNN is based on a backpropagation with time. However, the main challenge of the vanishing gradients is usually a reason for the parameters to take short-term dependencies while the information from earlier time-steps decays. The reverse issue, exploding gradients may be a reason for occurring the error to develop drastically with each time step.

LSTM is applied to the real dataset in a deep learning layer. An LSTM layer learns long-term dependencies between time steps in time-series and sequence data. The layer performs additive interactions, which can support improving gradient flow over long sequences during training.

To forecast the values of future time steps of a sequence, the training of the sequenced data regression based on the LSTM network. For each time step of the input sequence, the LSTM network learns to forecast the estimation value of the next time step.

### c. Results

The flowchart in Figure (4) is applied in this experiment for predicting the emergency cases of patients. Figure (6) presents a sample of predictive analysis for future data and time about 700.000 records concerning historical data. It applies the LSTM. The predictive analytics output includes continuous variables that are entitled a regression in Figure (7). An LSTM regression network was trained with 150 epochs; the LSTM layer contained 100 hidden units.

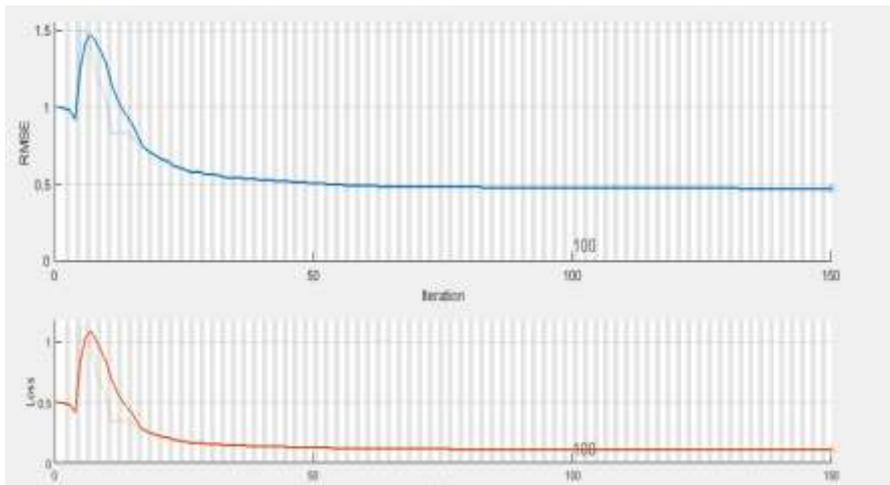

Figure 6: LSTM neural network training with 100 hidden layers with 150 epochs

This technique avoids the gradient explosion problem that is happening in artificial neural networks training data and backpropagation. The gradient condition was set to 1. The initial learning rate was set to 0.005 and then minimized after 150 epochs by multiplying by 0.2. LSTM network was trained with the specified training choices. Then the real measurements of the time proceedings between predictions are recognized, then the network state is updated with the observed values in the state the predicted values.

Resetting the network state prohibits prior predictions from affecting predictions on the new data. Thus, the network state is reset and then initialized by forecasting the training data. Predictions are made for each time step. For each forecast, the following time step values are predicted based on the observed values at the prior time step.

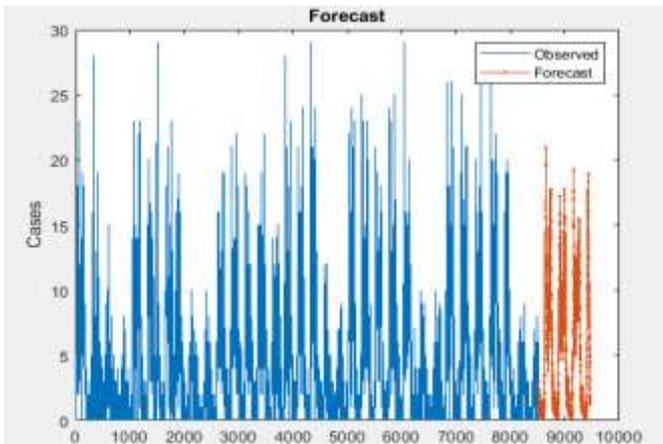

Figure.7: The predictive analysis of the data into a row vector

The 'Execution-Environment' option was set to 'predict-And-Update-State' using the 'CPU'. Non-standardized predictions were made utilizing the parameters evaluated earlier; then, the root-mean-square error (RMSE) was computed (as shown in equation 1).

The results of prediction achieve greater accuracy when the network state is edited with the notified values rather than the predicted values. Update Network State with Observed Values: When the existing values of the time steps between predictions are available, then the network state can be updated with the observed values instead of the predicted estimations. Then, the predicted values are compared with the test data. Figure (8) illustrates a comparison between the forecasted values and the test data. Figure (9) examines the forecasting with updates with RMSE =1.3.

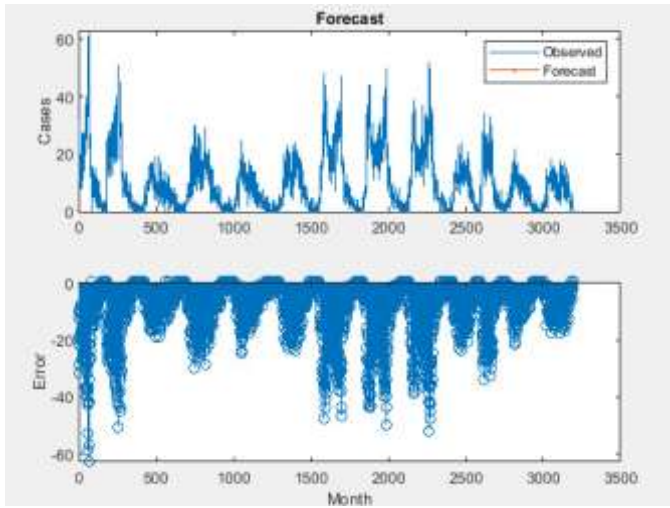

Figure.8: A Comparison between the forecasted values and the test data.

The x-axis that has been processed what the target shows that refers to the input data, so the constructed neural network has only one step of input data and it has seen 100 times and it has seen 500 steps of input data.

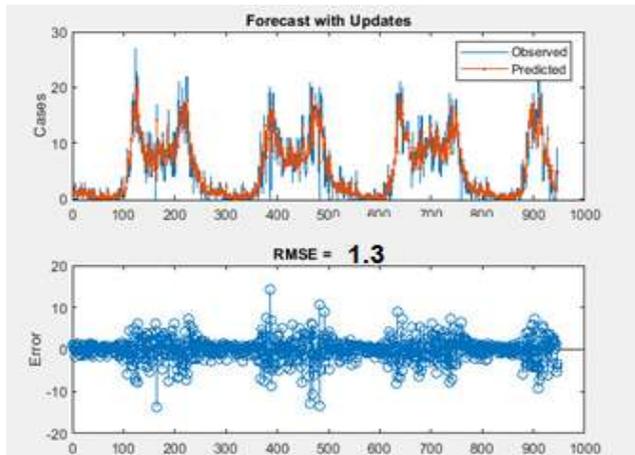
Figure.9: Forecast with updates

The summarized dataset interprets the results of number rates into ranges rates to make tracing easier. These results include risk levels (high, medium, low). These level drives doctors to make a suitable decision quickly and simultaneously (as shown in Table (6) and (7)). For examples, the patients reach to the high-risk level must go to the hospital if it is due the respiratory rate is high that must put them on a ventilator and requires making lung CT, if it is due the blood PH level that requires to take medicine and follow the Blood acidity if it is due the heart rate or body temperature, or blood pressure, taking medicine with Antiviral or Anti-malaria with Antipyretic.

Table.6: Tracing the risk level of the coronavirus's patients' cases

| Respiratory rate | Blood PH level | Heart rate | Blood pressure | Body Temperature | Risk level |
|---|---|---|---|---|---|
| Low | Normal | Normal | Normal | High | High |
| Lowest | Normal | Normal | Normal | Highest | High |
| Medium | Normal | Normal | Normal | Medium | Medium |
| High | Normal | Normal | Normal | Normal | Low |
| Normal | High | High | High | High | High |
| Normal | High | Normal | High | Medium | High |
| High | High | Normal | High | Highest | Highest |
| Medium | Medium | Medium | Medium | Medium | Medium |
| Medium | Medium | Low | High | Normal | Medium |
| High | Normal | Medium | Medium | Normal | Medium |
| low | Low | Normal | Low | High | High |
| lowest | Lowest | High | Low | Highest | Highest |
| Normal | Lowest | High | Normal | High | High |
| Normal | Normal | Normal | High | Medium | Medium |
| Normal | Normal | Medium | lowest | Normal | Medium |

The proposed system targets two essential objectives, monitoring infected patients remotely that are not high risk to avoid reaching high risk and predicting the next level of risk of each patient to protect their levels and taking decisions simultaneously.

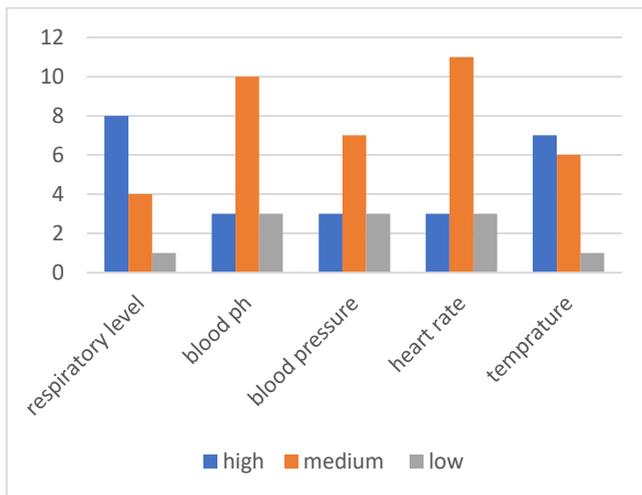

Figure.9: the explanation of three risk levels for the five parameters for coronavirus patients.

Table.7: the decisions for patient risk level

| Patients Risk level | Reasons | Decision |
|---|---|---|
| High-risk level | Respiratory rate | Must put a patient on ventilator and requires making lung CT |
| | Blood PH level | Should to Take medicine and follow the Blood acidity |
| | Heart rate | Taking medicine with Antiviral or Anti-malaria |
| | Blood pressure | Taking medicine with Antiviral or Anti-malaria |
| | Body Temperature | Taking Antipyretic |
| Medium-risk level | Respiratory rate | Must record to avoid reaching the high-risk level |
| | Blood PH level | Follow Acidity |
| | Heart rate | Taking medicine |
| | Blood pressure | Taking medicine |
| | Body Temperature | Taking Antipyretic and observing it continuously |
| Low-Risk level | Respiratory rate | Reassured case |
| | Blood PH level | Stable case |
| | Heart rate | Follow the heart pulse rate |
| | Blood pressure | Follow the stable case |
| | Body Temperature | Follow the stable case |

## VI. CONCLUSION AND FUTURE WORK

This paper presents the E-Quarantine system for monitoring infected patients with coronavirus remotely that uses for reducing infection and save hospital's places and equipment for high-risk patients only. The essential objectives of the E-Quarantine system that simulates the Quarantine for patients in their houses to monitor patients and classify the patients based on observing disease risks. The proposed system E-Quarantine monitors the patient's case flow and predicts the emergency cases around 24 hours by 98.7% based on the supervised previous data. It is based on five parameters, blood PH level, heart rate, blood pressure, body temperature, and respiratory rate. The proposed hybrid fusion is based on a hybrid of feature fusion level and decision fusion level that improves the accuracy results that reach 98.7%. Finding a Dempster-Shafer technique is more powerful in sequenced data than images or videos. The fusion technique is applied to sequenced data for patients and their respiratory sounds. For future work, the proposed system requires higher flexibility to be adaptive with multiple data types to improve the results of each patient.

**Compliance with Ethical Standards**: On behalf of all authors, the corresponding author states that there is no conflict of interest. This article does not contain any studies with human participants or animals performed by any of the authors.